\documentclass[fleqn,usenatbib]{rasti}

\usepackage{newtxtext,newtxmath}
\usepackage[T1]{fontenc}

\DeclareRobustCommand{\VAN}[3]{#2}
\let\VANthebibliography\thebibliography
\def\thebibliography{\DeclareRobustCommand{\VAN}[3]{##3}\VANthebibliography}

\usepackage{graphicx}	% Including figure files
\usepackage{amsmath}	% Advanced maths commands
\usepackage{subcaption}
\usepackage{makecell}

\usepackage{hyperref}
\usepackage{orcidlink}

\usepackage{threeparttable}
\usepackage{booktabs}

\title[Observing brown dwarfs with Ariel]{Assessing Ariel's capabilities to observe free-floating brown dwarfs}

\author[R. Akhmetshyn et al.]{
Roman Akhmetshyn \orcidlink{0009-0001-8160-162X},$^{1}$\thanks{E-mail: roman.akhmetshyn@mail.mcgill.ca}
Nicolas B. Cowan \orcidlink{0000-0001-6129-5699},$^{1,2}$
and Sarah Casewell \orcidlink{0000-0003-2478-0120} $^{3}$
\\
$^{1}$Department of Physics, McGill University, Montreal, Canada\\
$^{2}$Department of Earth \& Planetary Sciences, McGill University, Montreal, Canada\\
$^{3}$School of Physics and Astronomy, University of Leicester, University Road, Leicester LE1 7RH, UK
}

\date{Accepted XXX. Received YYY; in original form ZZZ}

\pubyear{\the\year{}}

\begin{document}
\label{firstpage}
\pagerange{\pageref{firstpage}--\pageref{lastpage}}
\maketitle

% Abstract of the paper
\begin{abstract}
The primary goal of the Ariel space telescope is to conduct the biggest spectroscopic survey of transiting exoplanets to characterize their atmospheres and weather. We propose to extend the Ariel survey to another domain of alien atmospheres – rogue planets and free-floating brown dwarfs. Their isolated nature means the observations are uncontaminated by light from a host star, and their short rotation periods, often similar to hot Jupiter orbital periods, provide an opportunity to study time-varying meteorology.  Phase curve observations would especially help scientists understand atmospheric dynamics at the L/T transition, where multiple cloud species at different altitudes influence the time-varying spectra of brown dwarfs. Inferring timescales and length scales of these atmospheric features is key to understanding the meteorology of sub-stellar objects. We quantify how many isolated cool objects that Ariel's fine guidance sensor (FGS) is able to guide on. Among 2744 selected targets, none are bright enough under the planned 10 Hz FGS cadence; however, with a "slow" fine guidance mode of 1 Hz, Ariel could study 98 L0- to L5-type brown dwarfs. We simulate single-epoch and time-series spectroscopic observations of the brightest isolated brown dwarfs given currently known instrumental specifications. We show that the resolution and sensitivity of Ariel instruments in the 1.1–7.8 micron regime can measure cloud-induced variability at the sub-percent level. A survey of brown dwarf phase curve observations, unavailable to ground-based telescopes, would be the perfect complement to Ariel's survey of atmospheric variability in hot Jupiters.

\end{abstract}

% Include between one and six keywords.

%%%%%%%%%%%%%%%%%%%%%%%%%%%%%%%%%%%%%%%%%%%%%%%%%%

%%%%%%%%%%%%%%%%% BODY OF PAPER %%%%%%%%%%%%%%%%%%

\section{Introduction}

\label{sec:intro}

Brown dwarfs and isolated planetary-mass objects (hereafter PMOs) are known to be photometrically and spectroscopically variable. Such behavior was theorized shortly after the first discovery of brown dwarfs \citep{tin99}, and then confirmed for objects across all late spectral types with dedicated ground- \citep{ bailer99, gelino2002dwarf, enoch2003photometric, morales2006sensitive, schmidt2007activity, wilson2014brown, vos2019search, liu2024near} and space-based observations \citep{buenzli2014brown, metchev2015weather, apai2017zones}. This variability arises from atmospheric inhomogeneities coupled with rotational modulation. The inhomogeneities encompass various phenomena: flaring from elevated magnetospheric star-like activity in earlier spectral types \citep{schmidt2007activity}, temperature fluctuations driven by convective cells and gravity waves \citep{robinson2014temperature}, radiative localized hot or cold spots,  vertical transport due to convective overshoot and chemical instability \citep{fegley1996atmospheric, griffith1999disequilibrium, freytag2010role}, eddy circulation in the stratosphere \citep{zhang2014atmospheric}, dust settling and heterogeneous cloud structures. The latter is especially important for the large-amplitude variability of objects in the L/T transition \citep{radigan2014strong}. This transition spans L7 to T4 spectral types and marks the condensation of multiple mineral species in the atmosphere at different opacity levels due to the gradual decrease in temperature with the age of a brown dwarf. For a review of brown dwarf variability, see \cite{artigau2018variability}.

Apart from rotational variability, brown dwarfs and PMOs display lightcurve \emph{evolution} from one rotation to the next due to the formation and breakup of clouds, differential rotation of these clouds at different latitudes, and migration of hot/cold spots. Measuring timescales of atmospheric variability can help uncover the physics governing brown dwarf meteorology. It is also important to determine how these timescales depend on the physical parameters of an object: its rotation velocity, age, surface gravity, metallicity, and magnetic field strength. 

Temporal atmospheric variability is likewise present on Hot Jupiters (HJs) and directly imaged exoplanets (DIEs). HJ variability has been predicted \citep{rauscher2007hot, komacek2020temporal, cho2021storms}, and tested by multiple Spitzer phase-curve observations \citep{agol2010climate, knutson2011spitzer, bell2019mass}. A synergy between Ariel and the James Webb Space Telescope to study HJ variability was also proposed and analyzed in \cite{changeat2025synergetic}. While young DIEs may differ from brown dwarfs in surface gravity and formation mechanism, their atmospheric physics are very similar, thus inhomogeneous cloud cover and thermochemical instabilities would produce quasi-periodic signals \citep{kostov2013mapping, biller2018exoplanet, sutlieff2023measuring}. Therefore, learning the meteorology of isolated substellar objects would directly aid studies of HJs and especially young DEIs.

Several observations confirmed the lightcurve evolution of brown dwarfs and PMOs in different photometric bands \citep[e.g.,][]{artigau2009photometric, gillon2013fast, girardin2013search, apai2021tess, brooks2023long}. Ground-based operations are interruptible and limited by seeing conditions; hence, only night-to-night changes in the lightcurve morphology for highly variable fast rotating objects ($<$4 hr) can be detected. The end product of such observations is a set of stop motion pictures that are unable to show the gradual change in weather patterns. Long-term, continuous, high-precision monitoring of the evolving atmospheres is only achievable with space telescopes. The proof of concept is the TESS observations of the Luhman 16 system covering over 540 hours \citep{apai2021tess}. This study was able to disentangle periodic signals from both components of the brown dwarf binary and found an evolution in rotation-induced variability explained by high-speed jets, zonal circulation, and planetary-scale waves; long-period evolution in the lightcurve was attributed to vortex-dominated polar regions. Although TESS has provided incredible lightcurve data on our closest brown dwarf neighbor, its optical photometric band and large pixel size hinder the detection of sub-percent variability in the rest of the brown dwarf population. 

A wide spectral range  is another important aspect of the meteorological study of brown dwarfs. The observed flux arises from different pressure levels at a given wavelength \citep{burrows1997nongray}. The lightcurve morphology and variability amplitude varies with wavelength, as different altitudes in the brown dwarf atmosphere have different meteorological processes, physical, and chemical conditions. As a result, time-series spectroscopic observations will encompass evolving signals from a variety of variability sources, and probe if they are correlated. 

In this paper, we assess the expansion of the Ariel space telescope's scope of operation to include a complementary survey of brown dwarf variability. Ariel will be the next major space instrument with an infrared spectrograph and wide wavelength coverage after the JWST. Recent publications and discoveries using JWST data show the power of space-based infrared facilities to study the weather on brown dwarfs with an unprecedented level of detail \citep{biller2024jwst, chen2025jwst, akhmetshyn2025mapping, mccarthy2025jwst, nasedkin2025jwst}. In Section \ref{sec:targets}, we describe the large sample of brown dwarfs we consider for the analysis. In Section \ref{sec:FGS}, we assess which targets can be guided given the nominal and enhanced guiding parameters of Ariel. In Section \ref{sec:spectra}, we simulate spectral observations of brown dwarfs with Ariel instruments. In Section \ref{sec:var}, we calculate the limiting variability magnitude and simulate spectrophotometric variability of brown dwarfs. 

\section{Target selection}
\label{sec:targets}

We build a target list from the Montreal Open Clusters and Associations Database (MOCAdb; \citealt{MOCAdb}) in combination with the UltracoolSheet database (DOI: 10.5281/zenodo.4169084). First and foremost, we filter out most of the ultra-cool main-sequence stars and young brown dwarfs; the latter are particularly challenging to correctly match with existing spectral models due to their reddened colors and interstellar extinction. Moreover, newly formed brown dwarfs have not contracted and spun up their rotation rate: observations show that the rotation periods of such objects are usually on the order of 1 day \citep{moore2019rotation} --- this makes variability characterization very time-demanding for them. We select objects of spectral type L0 and later but remove L0-L3 stars older than 1 Gyr old, as they likely belong to the main sequence \citep{baraffe2015new}. Using the available distance measurements and 2MASS J band photometry, we calculate the absolute magnitudes and remove objects brighter than M$_{\rm J}$=9, which are likely newly formed brown dwarfs or mismatched M-type stars. Objects without 2MASS photometry are kept in the analysis. 

\begin{figure}
    \centering
    \includegraphics[width=\linewidth]{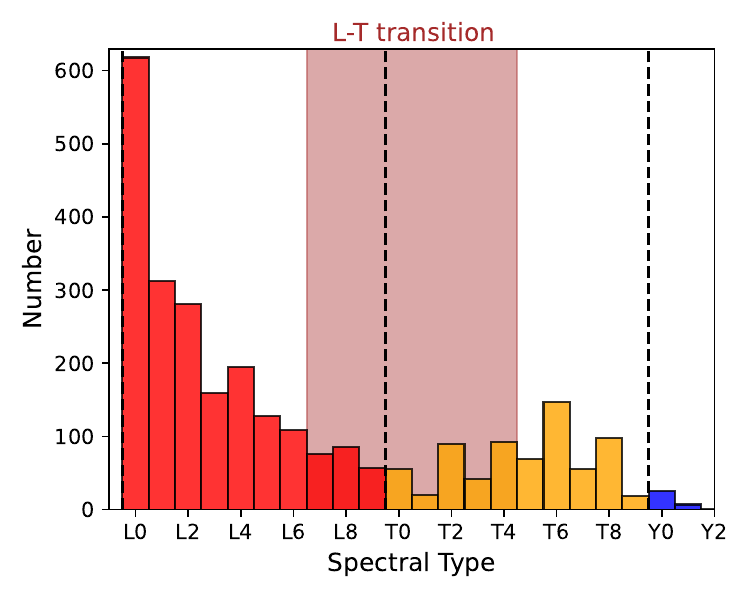}
    \caption{Distribution of brown dwarf spectral types across our sample. All objects have been assigned a spectral type based on their color, temperature, or observed spectrum. Spectral types with a fractional subtype have been rounded for this plot (e.g., T2.5 is equal to T3). L dwarfs dominate the sample as they are the brightest brown dwarfs. The decrease in sample count across the L-T transition is due to it being a time-constrained evolutionary transition: low-mass brown dwarfs cool faster and do not stay long in the L-T transition zone.}
    \label{fig:ST}
\end{figure}

In this study, we focus on isolated brown dwarfs and PMO's for uncontaminated benchmark observations of evolving weather. To ensure that our targets are isolated, we filter out objects that have a Gaia-detected neighbor within 3$\arcsec$, since the nominal slit width of Ariel instruments is 2.27$\arcsec$.  The final sample of 2744 potential targets spans the full sequence of L- and T-spectral types, and some Y-type objects (see Figure \ref{fig:ST}). Among these, 494 are in the L-T transition zone: spectral types L6.5 to T4.5. A color-magnitude diagram of the resulting sample is shown in Figure \ref{fig:CM}.

\section{Guiding capabilities}
\label{sec:FGS}

The feasibility of this proposed survey lies in the ability of Ariel to guide on objects from our sample. A fine guidance sensor (FGS) onboard the spacecraft is used to take sky images at 10 Hz cadence and maintain guiding. The second channel of the FGS operates in the 0.8 -- 1.1 $\mu$m range and would be most suitable for substellar objects that peak in the near-infrared. 

\begin{figure}
    \centering
    \includegraphics[width=0.95\linewidth]{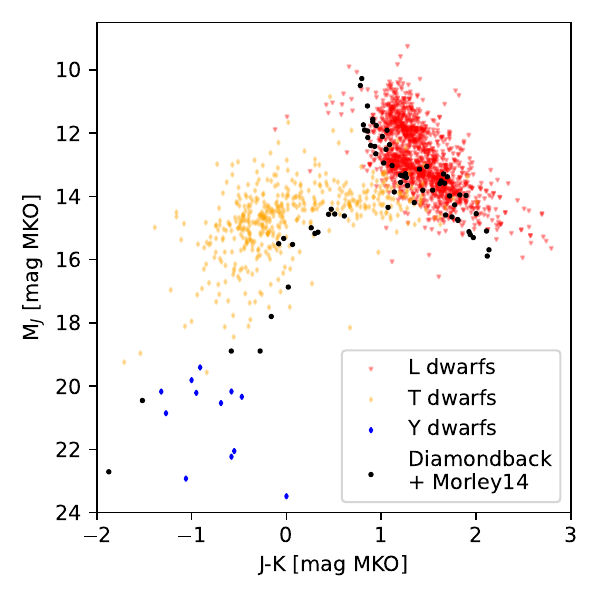}
    \caption{Color-magnitude diagram of 1905 out of 2744 potential Ariel brown dwarf targets with parallax measurements and MKO J and K photometric magnitudes.  Error bars are not shown. In addition to the large diversity in physical parameters, the scatter of data is attributed to large errors in distance estimations and the projection effect that influences the observed color for different inclinations \citep{vos2017viewing}. Black points represent NIR colors of the matched atmospheric models \citep{morley2014water, morley2024sonora}. The brown dwarf sequence is matched well, including the L-T transition.}
    \label{fig:CM}
\end{figure}

We estimate the photon flux at FGS2 wavelengths from all targets in the sample. The algorithm is as follows:
\begin{enumerate}
    \item We Match each brown dwarf with a spectral grid model using temperature, surface gravity, and metallicity from the catalog. For objects between 900 and 2400 K, we use Sonora family Diamondback spectral models \citep{morley2024sonora} that include a sedimentation efficiency parameter. We adopt an approximate f$_{\rm sed}$ to spectral type relation from \cite{saumon2008evolution}. Outside of that temperature range, we use grid models from \cite{morley2014water} (Morley14 hereafter). For objects without cataloged physical properties, we assume solar metallicity, typical log(g), and temperature for the spectral type and age. The accuracy of this matching can be evaluated from Figure \ref{fig:CM}, where we overplot NIR colors of the models on top of the color-magnitude diagram of our sample.
    \item We rescale the models, which are evaluated at the photosphere, using known K-band Vega magnitudes. Our targets have photometric K-band measurements from different facilities, so appropriate bandpasses are applied.
    \item We integrate the photon flux of rescaled models in the FGS2 wavelength bin: 0.8--1.1 $\mu$m.
\end{enumerate}

\begin{figure}
    \centering
    \includegraphics[width=\linewidth]{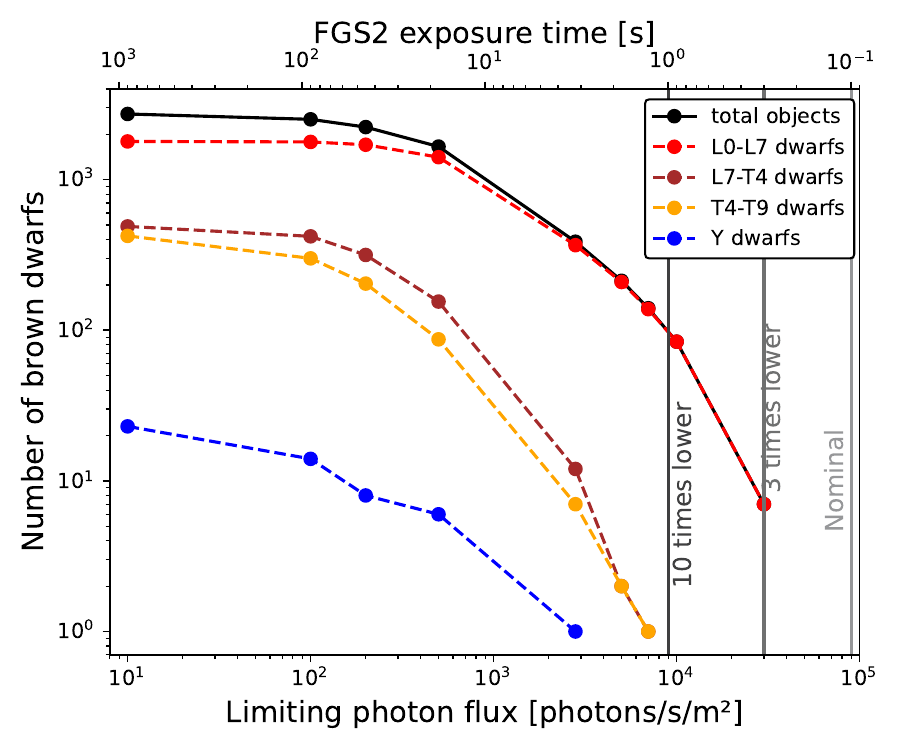}
    \caption{Cumulative number of observable brown dwarfs of different spectral types as a function of the limiting photon flux of the fine guidance sensor (FGS2).}
    \label{fig:guiding}
\end{figure}

The nominal sensitivity of the FGS at 10 Hz is $9\times10^4$ photons/s/m$^2$ (Billy Edwards, private communication). Given the effective collecting area of Ariel's primary mirror of 0.6 m$^2$, the FGS must be collecting at least 5400 photons per exposure. The cumulative number of guidable targets as a function of limiting photon flux is plotted in Figure \ref{fig:guiding}. There is no brown dwarf bright enough for the nominal threshold, but seven could be potentially guided on if the fine-guidance threshold was reduced by a factor of 3, a cadence of 3.3 Hz. A factor of 10 reduction in the signal threshold would allow Ariel to guide on 98 pre-L/T-transition brown dwarfs, but would decrease the FGS cadence to only 1 exposure per second. Such a "slow" fine guidance mode, still under consideration within the Ariel consortium, would allow fainter targets to be observed and greatly benefit brown dwarf and exoplanet science alike, as more M-dwarf planets could be surveyed.

\section{Spectral observations}
\label{sec:spectra}

Spectroscopic observations of brown dwarfs can tell us a lot about their physical properties and atmospheric composition. Broad wavelength coverage provides the tightest abundance constraints on bolometric luminosity and atmospheric retrievals \citep{burningham2021cloud}. Space-based spectroscopy provides high signal-to-noise (SNR) data without the need for telluric correction. 

%Even though Ariel has low-resolution spectrographs, \cite{mugnai2021alfnoor} \cite{edwards2022ariel}, and \cite{bocchieri2023detecting} showed that the presence of certain molecules in the transmission spectra of exoplanets can certainly be detected and constrained, even for Tier 1 observations, which only provide 10 data points for  0.5–7.8 $\mu$m wavelength range. Observations of brown dwarfs, unlike transiting exoplanets, already provide high SNR over a wide spectral range without the need for binning. The resolution of 3 instruments is sufficient to capture broadband molecular features that shape the atmospheric spectrum of isolated brown dwarfs.

Using the matched atmospheric models in the previous section, we calculate Poisson noise for unbinned spectroscopic observations using Ariel's three spectrograph instruments, with 1 minute exposures for all targets, accounting for the telescope's effective collecting area and average transmission ($\tau$) of each instrument:
\begin{itemize}
    \item NIRSpec ($\lambda \in[1.1,1.95]$ $\mu$m, $R$ = 20, $\tau$ = 0.27);
 \item AIRS CH0 ($\lambda \in[1.95,3.9]$ $\mu$m, $R$ = 100, $\tau$ = 0.18);
 \item AIRS CH1 ($\lambda \in[3.9, 7.8]$ $\mu$m, $R$ = 30, $\tau$ = 0.18).
\end{itemize}

\begin{figure*}
    \centering
    \includegraphics[width=\linewidth]{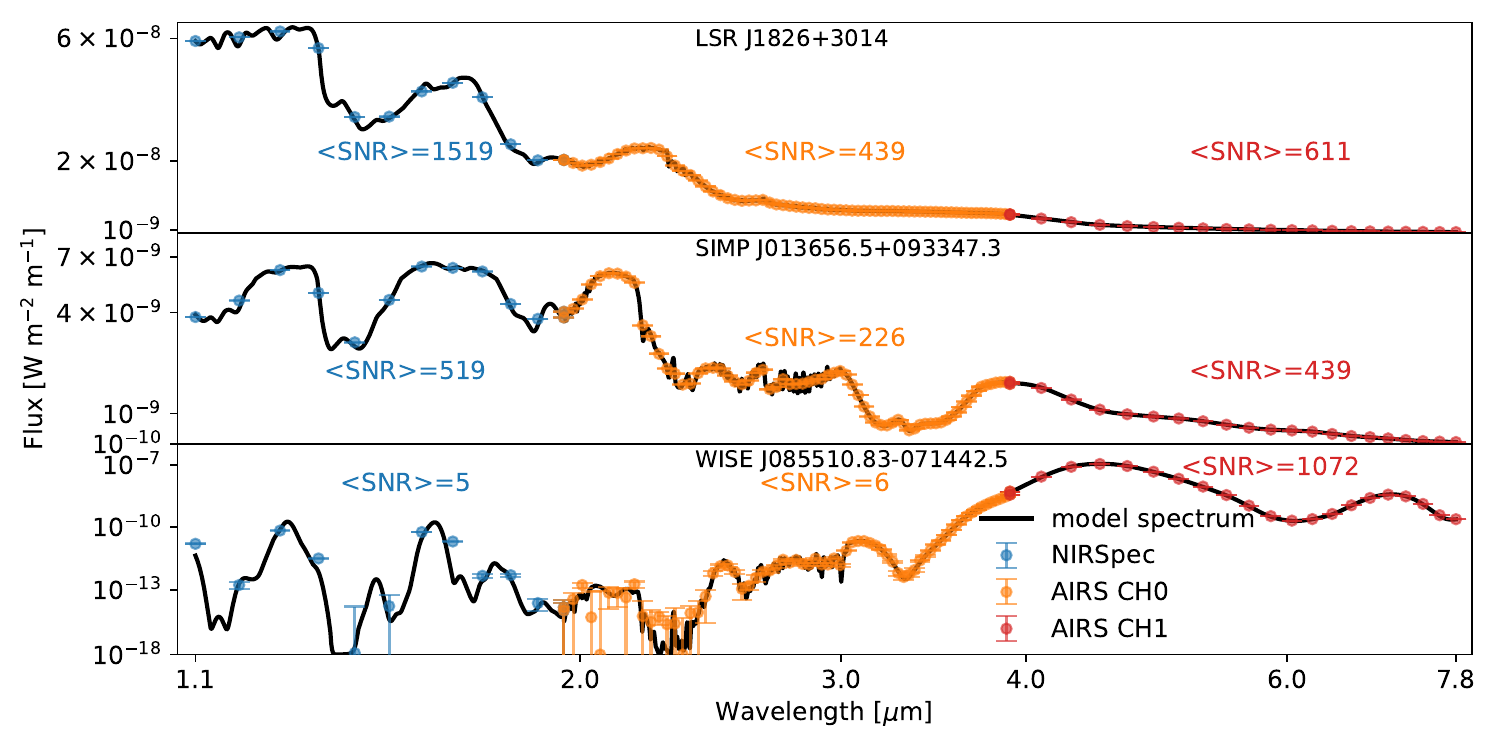}
    \caption{Simulated Ariel single-epoch spectral observations of 3 targets from our sample. Each target was observed with a 1-minute exposure --- enough to capture rotational variability. The integration time could safely be increased to 5 minutes for higher SNR on fainter targets. \emph{Top}: LSR J1826+3014, the brightest free-floating brown dwarf in the FGS2 bandpass, an L0 dwarf. \emph{Middle}: SIMP J0136, a well-known variable T2.5 dwarf at the L/T transition. \emph{Bottom}: WISE 0855-0714, the closest known Y dwarf. These targets emit 5.5$\times$10$^4$, 0.36$\times$10$^4$, and  270 photons/s/m$^2$ respectively in the FGS2 band. Despite the high expected SNR with Ariel's spectroscopic instruments, guiding on SIMP J0136 would require 2.5-second FGS2 readout (33.3 seconds for the Y dwarf). Additional analysis of the spacecraft stability and centroiding accuracy will be required to assess how much the guiding cadence can be decreased. SNRs reported on the plot are median values for each instrument. Black lines are smoothed spectra of corresponding theoretical models: Sonora Diamondback \citep{morley2024sonora} for  LSR J1826 and SIMP J0136, and Morley14 \citep{morley2014water} for WISE 0855.}
    \label{fig:spectra}
\end{figure*}

In Figure \ref{fig:spectra}, we show the simulated Ariel observations of three targets: a bright L dwarf, a brown dwarf at the L-T transition, and the nearest Y dwarf. Our simulations predict high SNR across all instruments for the majority of sample targets, with the exception of Y dwarfs and late T dwarfs, yet the former have the highest SNRs in AIRS CH1 due to the W2 band opacity window (3.9 -- 5.3 $\mu$m). 
%Unlike ground-based surveys, the noise will be limited strictly by photon noise, detector gain, readout noise, dark current, and Poisson noise from thermal and zodiacal emission. 
Estimates of instrumental sources of noise are not yet publicly available, as the telescope is still under construction and testing, but photon noise will likely be the dominant source of noise for most observable targets. 
%We predict that the SNR could be increased to our predicted values by binning in wavelength and co-adding exposures in up to 5-minute bins, which will not compromise the sensitivity to rotation-driven spectrophotometric variability.

\section{Spectrophotometric variability}
\label{sec:var}
Although single-epoch spectroscopic observations are useful for constraining spectral types and physical properties of a large sample of brown dwarfs, the key idea of our proposed survey is to analyze time-dependent spectrophotometric variability. Broad wavelength coverage of this variability is especially important because emergent flux at different wavelengths probes
different depths in the atmospheres \citep{burrows1997nongray, yang2016extrasolar}. %As different meteorological processes occur and various molecules condense into clouds at those depths, we expect to observe different lightcurve morphology and phase shifts between wavelength bins of 3 instruments. 
For example, flux emerging from below the cloud base will be modulated by the spatial distribution of those clouds across the globe, while fluxes from above the cloud tops are governed by localized heating and abundances. By observing lightcurve variability at different wavelength bands, we can infer a 3-dimensional (longitude, latitude, and depth) map of an evolving atmosphere \citep{plummer2026mapping}. 

\begin{figure*}
    \centering
    \includegraphics[width=1\linewidth]{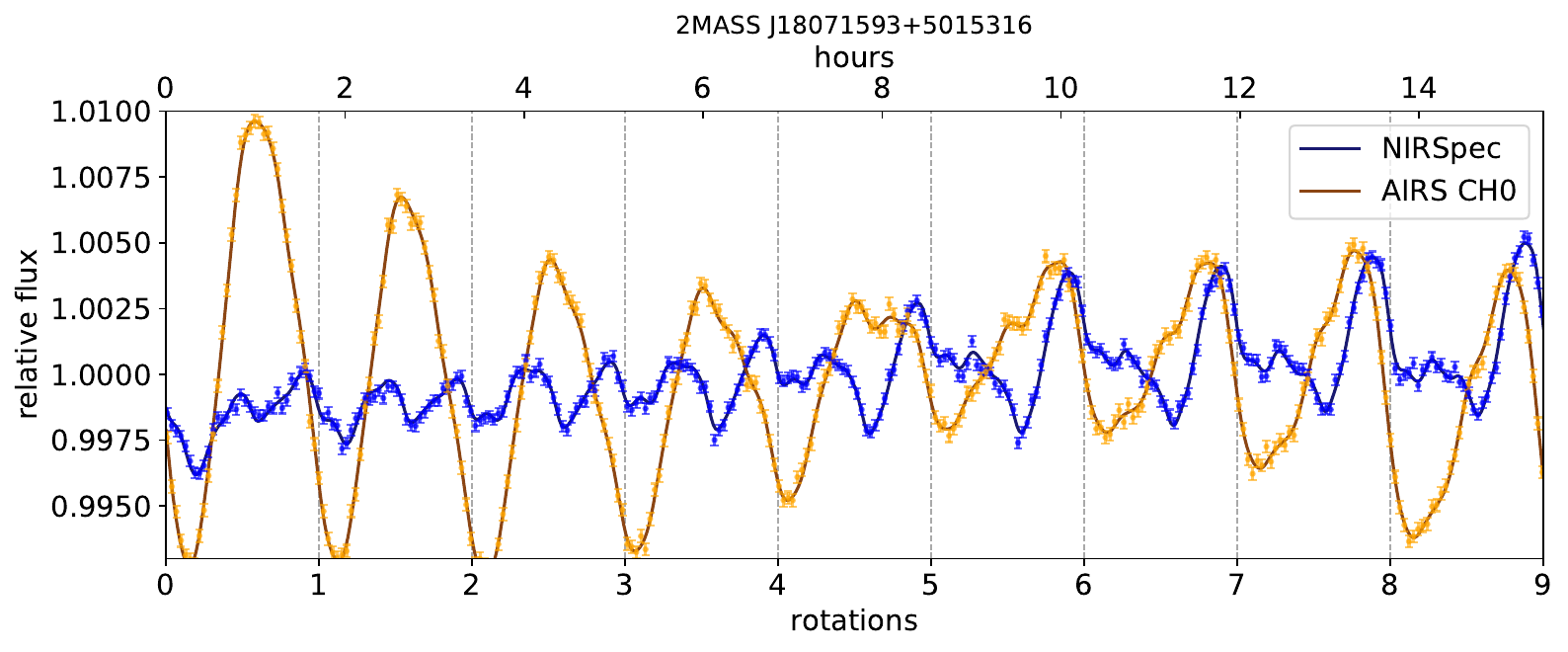}
    \caption{Simulated photometric observations in Ariel's NIRSpec and AIRS CH0 bands of an L1.5-type brown dwarf 2MASS J18071593+5015316 with a known rotation period of 1.71 hours \citep{miles2017rotation}. The simulated lightcurves (solid lines) were generated with \texttt{Starry-Process} \citep{starry_p}. Gaussian noise was added with standard deviation corresponding to the SNR. The simulated data show lightcurves binned across whole instrument wavelength range and in 3-minute windows.}
    \label{fig:lc_evo}
\end{figure*}

To estimate the detectability of brown dwarf variability, we simply calculate the median relative error for each instrument for each target, using the same model spectra integrated for 1 minute as in the previous section. This value represents the limit below which variability on minute timescales becomes indistinguishable from photon noise. The SNR can be improved even further by binning in wavelength, especially in the AIRS CH0 band. 
We show simulated Ariel multi-band lighturve observations for one of the brightest targets with a known rotational period in Figure \ref{fig:lc_evo}. The lightcurve morphologies across wavelength bands and their evolution are based on previous observations of brown dwarfs \citep{croll2016long, akhmetshyn2025mapping}. We predict that sub-percent variability would be detected at one-minute exposures for all 98 early L-type brown dwarfs that could be guided on at 1 Hz cadence (listed in Table \ref{tab:best}). Observations in the NIRSpec band could even detect variability below 0.1\% amplitude  for all these targets. Naturally, if the guiding flux was not a limiting factor, sub-percent variability could be observed across a greater number of isolated brown dwarfs from our sample. 

%\begin{figure}
%    \centering
%    \includegraphics[width=0.95\linewidth]{mag_var.pdf}
%    \caption{Estimated photon flux in the FGS2 wavelength bin and the lowest detectable variability magnitude of all targets in our sample. Grey lines indicate 1\% variability and 3$\times$10$^4$ photons/s/m$^2$ limiting flux in the FGS2 band. Cyan line indicates the lowest photon flux required to observe all objects with sub-percent variability for the given instrument. The most favorable targets for observations are in the lower right corner and represent early-type brown dwarfs (T0--T5).}
%    \label{fig:varmag}
%\end{figure}

Rotational period is a less common measurement for brown dwarfs, as it requires high photometric stability and sensitivity. In our sample, only 72 objects have rotation period estimations, most of which are derived from the Hubble Space Telescope Weather on Other Worlds survey results \citep{metchev2015weather, tannock2021weather}. Almost all isolated brown dwarfs are fast rotators with spin periods $\leqslant$24 hours \citep{scholz2015rotation, moore2019rotation}, with an exception for very young objects in the contraction phase. Within our sample, 64\% have periods $\leqslant$5 hours -- comparable to the duration of a typical transit observation.  

Given this information, it would be safe to dedicate 10-hour-long time-series observations for all targets with unknown rotation periods. It will be enough to determine whether an object is a fast rotator ($P\leqslant$ 5 hours), cover multiple rotation phases, and detect lightcurve evolution due to evolving meteorology. Unlike fast rotators, objects with longer rotation periods display lightcurve evolution on longer timescales \citep[e.g.][]{radigan2012large}; these could be revisited further with more time dedicated to follow-up observations. Among the 98 brightest brown dwarfs in our sample, 12 have measured periods of less than five hours and 1 of nine hours. If we dedicate 5--10 rotation periods for each, observing these would require 222 -- 444 hours of science time. Observing all 98 targets for a blind periodicity survey (10 hr for each target) would require $\sim$2.8\% of Ariel's nominal mission time. For context, 5--10\% of the primary mission time will be dedicated to complementary science.

\section{Discussion and Conclusion}
\label{sec:discussion}

In this study, we assessed whether the Ariel space telescope could perform long-baseline observations of isolated PMO and brown dwarf variability.  Our final sample of 2744 objects includes 2032 L dwarfs, 687 T dwarfs, and 25 Y dwarfs. Despite our filtering approach applied in Section \ref{sec:targets}, some objects may still be M dwarf companions or brown dwarf binaries. The MOCAdb was recently released and is constantly being updated, as is the UltraCoolSheet, therefore some discrepancies in observed properties are to be expected, especially for the faint objects. 

By matching atmospheric models to each target and integrating the flux in specific wavelength bins, we determined that 97\% of the targets we considered for the analysis could provide SNR above 10 with only 1-minute exposures and no binning in wavelength, in the photon noise limit. However, none of these targets can be guided on, as the nominal flux required for the fine guidance sensor is too high. A potential "slow" guiding mode could greatly increase the number of guidable isolated substellar objects. With 1-second long FGS exposures, Ariel could observe about 98 L dwarfs, all of which belong to early-type (L0-L5). \cite{radigan2014independent} shows that most variable brown dwarfs are predictably those in the L-T transition, where cloud decks sink below the photosphere. Nevertheless, that study, along with \cite{metchev2015weather} and \cite{lew2016cloud}, also confirm 1\% to 4\% variability magnitude for early-type L dwarfs, based on observations in J, Spitzer IRAC Ch1, and IRAC Ch2 bands. This is promising, because our calculations show that for all these 98 potential targets, we can confidently detect sub-percent, in some cases even tenth of a percent, variability. 

Our long-baseline observing strategy of 10 hours per target is time-demanding, especially if all 98 targets were to be observed. A good target selection approach should be utilized that maximizes the statistical leverage \citep{cowan2025maximizing, panek2026balancing}, i.e., the diversity in physical parameters of the targets such as mass, surface gravity, temperature, age, and metallicity. Observing a single target of each specific property will not be viable due to the projection effects: bands that probe deeper layers of the atmosphere are more attenuated because the path lengths increase at lower inclinations \citep{vos2017viewing}.
%The same study also finds the evidence that the variability amplitude may increase with rotational period for periods up to 10 hours. This should not be discouraging from pursuing faster rotating brown dwarfs in the target selection, as Ariel's instruments will be sensitive to sub-percent variability commonly detected for these objects \citep{vos2017viewing}. 
It is likewise important to infer sources of weather variability for objects with a variety of rotation periods.  The whole list of 98 brightest targets and some of their known properties is included in Appendix \ref{sec:appendix1}.

Our proposed variability survey of brown dwarfs cannot be performed within the schedule gaps of Ariel's primary exoplanet survey. These schedule gaps are predicted to be mostly 1 to 2 hours long (Juan Carlos Morales, private communication). Given the spacecraft slew rate of 4.5$^\circ$/min, and approximate set-up time of 5 minutes (Matt Griffin, private communication), only 30 to 60 minutes of observing time is available for a typical gap. During these gaps, Ariel could instead perform single-epoch spectroscopic observations of isolated brown dwarfs at high SNR. The complementary survey of atmospheric variability could then be scheduled during Ariel's 2 year extended mission, when observing time could be specifically dedicated to a periodicity survey (10 hr each target), or meteorology survey (5 to 10 rotations for each target). Unlike exoplanetary observations, these are not time-sensitive, as long as there is a long-baseline and continuous monitoring. 

To summarize, the Ariel space telescope can perform a large complementary survey of atmospheric variability among isolated brown dwarfs and planetary mass objects. It is capable of almost uninterrupted long-baseline observations of weather-induced lightcurve evolution across a wide wavelength range at high SNR --- unattainable for ground-based facilities. %The results could be utilized to create three-dimensional evolving atmospheric maps using \texttt{Starry} \citep{luger2019starry} or \texttt{Jaxoplanet} \citep{jaxoplanet}; improve current models by performing atmospheric retrievals at different rotation phases; deepen our understanding of dynamic atmospheres; decipher what role does turbulence, winds and convection play on fast rotating objects. 
Since the atmospheres of substellar objects are similar in temperature and composition to young directly imaged exoplanets, the results of the survey would simultaneously enhance exoplanetary science. 

These ambitious goals are achievable only if the Ariel is capable of guiding on these faint objects. Our results show that by decreasing the cadence of the fine guiding sensor, we can increase the scientific output of the mission. Slower FGS readout would benefit not only brown dwarf science but also the study of planets around M-dwarfs. Further analysis of the centroiding accuracy and spacecraft's stability is required to determine the longest practical exposure times for the FGS correcting images.

\section*{Acknowledgements}

This work was supported by NSERC Discovery Grant, Canada Research Chair, and McDonald Fellowship. The authors also thank the Trottier Space Institute and l’Institut de recherche sur les exoplanétes for their financial support and collaborative environment, as well as the Center for research in astrophysics of Quebec (CRAQ)/AstroQuebec.

RA is grateful for Jonathan Gagn\'e's immense help with MOCAdb.

The authors also acknowledge the dynamic intellectual environment of the conference “Ariel: Science, Mission \& Community 2026” at which they received valuable feedback for this work.

This research made use of the Montreal Open Clusters and Associations (MOCA) database, operated at the Montr\'eal Plan\'etarium \citep{MOCAdb}.

This work has benefited from \href{http://bit.ly/UltracoolSheet}{The UltracoolSheet}, maintained by Will Best, Trent Dupuy, Michael Liu, Aniket Sanghi, Rob Siverd, and Zhoujian Zhang, and developed from compilations by \cite{dupuy2012hawaii, dupuy2013distances, deacon2014wide, liu2016hawaii, best2018photometry, best2021volume, sanghi2023hawaii, schneider2023astrometry}.	

\section*{Data availability}
The \href{http://bit.ly/UltracoolSheet}{The UltracoolSheet} online database of known brown dwarfs and the Montreal Open Clusters and Associations (MOCA) database that were used in this research are publicly available.

%%%%%%%%%%%%%%%%%%%% REFERENCES %%%%%%%%%%%%%%%%%%

% The best way to enter references is to use BibTeX:

\bibliographystyle{rasti}
\bibliography{refs} % if your bibtex file is called example.bib

@article{tin99,
    author = {Tinney, C. G. and Tolley, A. J.},
    title = {Searching for weather in brown dwarfs},
    journal = {Monthly Notices of the Royal Astronomical Society},
    volume = {304},
    number = {1},
    pages = {119-126},
    year = {1999},
    month = {03},
    issn = {0035-8711},
    doi = {10.1046/j.1365-8711.1999.02297.x},
    url = {https://doi.org/10.1046/j.1365-8711.1999.02297.x},
    eprint = {https://academic.oup.com/mnras/article-pdf/304/1/119/2931392/304-1-119.pdf},
}

@ARTICLE{bailer99,
       author = {{Bailer-Jones}, C.~A.~L. and {Mundt}, R.},
        title = "{A search for variability in brown dwarfs and L dwarfs}",
      journal = {\aap},
         year = 1999,
        month = aug,
       volume = {348},
        pages = {800-804},
          doi = {10.48550/arXiv.astro-ph/9906439},
archivePrefix = {arXiv},
       eprint = {astro-ph/9906439},
 primaryClass = {astro-ph},
       adsurl = {https://ui.adsabs.harvard.edu/abs/1999A&A...348..800B}
}

@article{gelino2002dwarf,
  title={L dwarf variability: I-band observations},
  author={Gelino, Christopher R and Marley, Mark S and Holtzman, Jon A and Ackerman, Andrew S and Lodders, Katharina},
  journal={The Astrophysical Journal},
  volume={577},
  number={1},
  pages={433--446},
  year={2002}
}

@article{enoch2003photometric,
  title={Photometric variability at the L/T dwarf boundary},
  author={Enoch, Melissa L and Brown, Michael E and Burgasser, Adam J},
  journal={The Astronomical Journal},
  volume={126},
  number={2},
  pages={1006--1016},
  year={2003}
}

@article{morales2006sensitive,
  title={A sensitive search for variability in late L dwarfs: the quest for weather},
  author={Morales-Calder{\'o}n, Maria and Stauffer, JR and Kirkpatrick, J Davy and Carey, S and Gelino, CR and Navascu{\'e}s, D Barrado y and Rebull, L and Lowrance, P and Marley, MS and Charbonneau, D and others},
  journal={The Astrophysical Journal},
  volume={653},
  number={2},
  pages={1454--1463},
  year={2006}
}

@article{radigan2014strong,
  title={Strong brightness variations signal cloudy-to-clear transition of brown dwarfs},
  author={Radigan, Jacqueline and Lafreni{\`e}re, David and Jayawardhana, Ray and Artigau, Etienne},
  journal={The Astrophysical Journal},
  volume={793},
  number={2},
  pages={75},
  year={2014},
  publisher={The American Astronomical Society}
}

@article{wilson2014brown,
  title={The brown dwarf atmosphere monitoring (BAM) project-I. The largest near-IR monitoring survey of L and T dwarfs},
  author={Wilson, PA and Rajan, A and Patience, Jennifer},
  journal={Astronomy \& Astrophysics},
  volume={566},
  pages={A111},
  year={2014},
  publisher={EDP Sciences}
}

@article{metchev2015weather,
  title={Weather on other worlds. II. Survey results: spots are ubiquitous on L and T dwarfs},
  author={Metchev, Stanimir A and Heinze, Aren and Apai, D{\'a}niel and Flateau, Davin and Radigan, Jacqueline and Burgasser, Adam and Marley, Mark S and Artigau, {\'E}tienne and Plavchan, Peter and Goldman, Bertrand},
  journal={The Astrophysical Journal},
  volume={799},
  number={2},
  pages={154},
  year={2015},
  publisher={The American Astronomical Society}
}

@article{buenzli2014brown,
  title={Brown dwarf photospheres are patchy: a hubble space telescope near-infrared spectroscopic survey finds frequent low-level variability},
  author={Buenzli, Esther and Apai, D{\'a}niel and Radigan, Jacqueline and Reid, I Neill and Flateau, Davin},
  journal={The Astrophysical Journal},
  volume={782},
  number={2},
  pages={77},
  year={2014},
  publisher={The American Astronomical Society}
}

@article{apai2017zones,
  title={Zones, spots, and planetary-scale waves beating in brown dwarf atmospheres},
  author={Apai, Daniel and Karalidi, Theodora and Marley, Mark S and Yang, Hao and Flateau, Davin and Metchev, Stanimir and Cowan, NB and Buenzli, Esther and Burgasser, Adam J and Radigan, Jacqueline and others},
  journal={Science},
  volume={357},
  number={6352},
  pages={683--687},
  year={2017},
  publisher={American Association for the Advancement of Science}
}

@article{liu2024near,
  title={A near-infrared variability survey of young planetary-mass objects},
  author={Liu, Pengyu and Biller, Beth A and Vos, Johanna M and Whiteford, Niall and Zhang, Zhoujian and Liu, Michael C and Fontanive, Cl{\'e}mence and Manjavacas, Elena and Henning, Thomas and Kenworthy, Matthew A and others},
  journal={Monthly Notices of the Royal Astronomical Society},
  volume={527},
  number={3},
  pages={6624--6674},
  year={2024},
  publisher={Oxford University Press}
}

@article{vos2019search,
  title={A search for variability in exoplanet analogues and low-gravity brown dwarfs},
  author={Vos, Johanna M and Biller, Beth A and Bonavita, Mariangela and Eriksson, Simon and Liu, Michael C and Best, William MJ and Metchev, Stanimir and Radigan, Jacqueline and Allers, Katelyn N and Janson, Markus and others},
  journal={Monthly Notices of the Royal Astronomical Society},
  volume={483},
  number={1},
  pages={480--502},
  year={2019},
  publisher={Oxford University Press}
}

@article{schmidt2007activity,
  title={Activity and kinematics of ultracool dwarfs, including an amazing flare observation},
  author={Schmidt, Sarah J and Cruz, Kelle L and Bongiorno, Bethany J and Liebert, James and Reid, I Neill},
  journal={The Astronomical Journal},
  volume={133},
  number={5},
  pages={2258--2273},
  year={2007}
}

@article{robinson2014temperature,
  title={Temperature fluctuations as a source of brown dwarf variability},
  author={Robinson, Tyler D and Marley, Mark S},
  journal={The Astrophysical Journal},
  volume={785},
  number={2},
  pages={158},
  year={2014},
  publisher={The American Astronomical Society}
}

@article{freytag2010role,
  title={The role of convection, overshoot, and gravity waves for the transport of dust in M dwarf and brown dwarf atmospheres},
  author={Freytag, Bernd and Allard, France and Ludwig, H-G and Homeier, Derek and Steffen, Matthias},
  journal={Astronomy \& Astrophysics},
  volume={513},
  pages={A19},
  year={2010},
  publisher={EDP Sciences}
}

@article{fegley1996atmospheric,
  title={Atmospheric chemistry of the brown dwarf Gliese 229B: Thermochemical equilibrium predictions},
  author={Fegley, Jr, Bruce and Lodders, Katharina},
  journal={The Astrophysical Journal Letters},
  volume={472},
  number={1},
  pages={L37--L39},
  year={1996}
}

@article{griffith1999disequilibrium,
  title={Disequilibrium chemistry in a brown dwarf's atmosphere: carbon monoxide in Gliese 229B},
  author={Griffith, Caitlin A and Yelle, Roger V},
  journal={The Astrophysical Journal Letters},
  volume={519},
  number={1},
  pages={L85--L88},
  year={1999}
}

@article{zhang2014atmospheric,
  title={Atmospheric circulation of brown dwarfs: jets, vortices, and time variability},
  author={Zhang, Xi and Showman, Adam P},
  journal={The Astrophysical Journal Letters},
  volume={788},
  number={1},
  pages={L6},
  year={2014},
  publisher={The American Astronomical Society}
}

@article{gillon2013fast,
  title={Fast-evolving weather for the coolest of our two new substellar neighbours},
  author={Gillon, Micha{\"e}l and Triaud, AHMJ and Jehin, Emmanuel and Delrez, Laetitia and Opitom, Cyrielle and Magain, Pierre and Lendl, M and Queloz, D},
  journal={Astronomy \& Astrophysics},
  volume={555},
  pages={L5},
  year={2013},
  publisher={EDP Sciences}
}

@article{artigau2009photometric,
  title={Photometric variability of the T2. 5 brown dwarf SIMP J013656. 5+ 093347: Evidence for evolving weather patterns},
  author={Artigau, {\'E}tienne and Bouchard, Sandie and Doyon, Ren{\'e} and Lafreni{\`e}re, David},
  journal={The Astrophysical Journal},
  volume={701},
  number={2},
  pages={1534--1539},
  year={2009},
  publisher={The American Astronomical Society}
}

@article{apai2021tess,
  title={TESS observations of the Luhman 16 AB Brown Dwarf system: Rotational periods, lightcurve evolution, and zonal circulation},
  author={Apai, D{\'a}niel and Nardiello, Domenico and Bedin, Luigi R},
  journal={The Astrophysical Journal},
  volume={906},
  number={1},
  pages={64},
  year={2021},
  publisher={The American Astronomical Society}
}

@article{girardin2013search,
  title={In search of dust clouds: photometric monitoring of a sample of late L and T dwarfs},
  author={Girardin, F and Artigau, {\'E} and Doyon, R},
  journal={The Astrophysical Journal},
  volume={767},
  number={1},
  pages={61},
  year={2013},
  publisher={The American Astronomical Society}
}

@article{brooks2023long,
  title={Long-term 4.6 $\mu$ m Variability in Brown Dwarfs and a New Technique for Identifying Brown Dwarf Binary Candidates},
  author={Brooks, Hunter and Kirkpatrick, J Davy and Meisner, Aaron M and Gelino, Christopher R and Bardalez Gagliuffi, Daniella C and Marocco, Federico and Schneider, Adam C and Faherty, Jacqueline K and Casewell, SL and Raghu, Yadukrishna and others},
  journal={The Astronomical Journal},
  volume={165},
  number={6},
  pages={232},
  year={2023},
  publisher={The American Astronomical Society}
}

@article{mccarthy2025jwst,
  title={The JWST weather report from the isolated exoplanet analog SIMP 0136+ 0933: Pressure-dependent variability driven by multiple mechanisms},
  author={McCarthy, Allison M and Vos, Johanna M and Muirhead, Philip S and Biller, Beth A and Morley, Caroline V and Faherty, Jacqueline and Burningham, Ben and Calamari, Emily and Cowan, Nicolas B and Cruz, Kelle L and others},
  journal={The Astrophysical Journal Letters},
  volume={981},
  number={2},
  pages={L22},
  year={2025},
  publisher={The American Astronomical Society}
}

@article{nasedkin2025jwst,
  title={The JWST weather report: Retrieving temperature variations, auroral heating, and static cloud coverage on SIMP-0136},
  author={Nasedkin, Evert and Schrader, Merle and Vos, Johanna M and Biller, Beth and Burningham, Ben and Cowan, Nicolas B and Faherty, JK and Gonzales, Eileen and Lam, Madeline B and McCarthy, Allison M and others},
  journal={Astronomy \& Astrophysics},
  volume={702},
  pages={A1},
  year={2025},
  publisher={EDP Sciences}
}

@article{akhmetshyn2025mapping,
  title={Mapping atmospheric features of the planetary-mass brown dwarf SIMP 0136 with JWST NIRISS},
  author={Akhmetshyn, Roman and Artigau, {\'E}tienne and Cowan, Nicolas B and Plummer, Michael K and Wang, Fei and Burningham, Ben and Benneke, Bj{\"o}rn and Doyon, Ren{\'e} and Jayawardhana, Ray and Lafreniere, David and others},
  journal={The Astrophysical Journal},
  volume={993},
  number={2},
  pages={237},
  year={2025},
  publisher={The American Astronomical Society}
}

@article{biller2024jwst,
  title={The JWST weather report from the nearest brown dwarfs I: multiperiod JWST NIRSpec+ MIRI monitoring of the benchmark binary brown dwarf WISE 1049AB},
  author={Biller, Beth A and Vos, Johanna M and Zhou, Yifan and McCarthy, Allison M and Tan, Xianyu and Crossfield, Ian JM and Whiteford, Niall and Suarez, Genaro and Faherty, Jacqueline and Manjavacas, Elena and others},
  journal={Monthly Notices of the Royal Astronomical Society},
  volume={532},
  number={2},
  pages={2207--2233},
  year={2024},
  publisher={Oxford University Press}
}

@article{chen2025jwst,
  title={The JWST weather report from the nearest brown dwarfs II: consistent variability mechanisms over 7 months revealed by 1--14 $\mu$m NIRSpec+ MIRI monitoring of WISE 1049AB},
  author={Chen, Xueqing and Biller, Beth A and Tan, Xianyu and Vos, Johanna M and Zhou, Yifan and Su{\'a}rez, Genaro and McCarthy, Allison M and Morley, Caroline V and Whiteford, Niall and Dupuy, Trent J and others},
  journal={Monthly Notices of the Royal Astronomical Society},
  volume={539},
  number={4},
  pages={3758--3777},
  year={2025},
  publisher={Oxford University Press}
}

@ARTICLE{MOCAdb,
       author = {{Gagn{\'e}}, Jonathan and {Moranta}, Leslie and {Faherty}, Jacqueline K. and {Curtis}, Jason Lee and {Bickle}, Thomas P. and {Couture}, Dominic and {Chiasson David}, Am{\'e}lie and {Christie}, Katie and {Lambier}, Samantha and {Leclerc}, Elise and {Poliquin}, Livia and {Belzile}, Danika and {Mamajek}, Eric E.},
        title = "{The Montreal Open Clusters and Associations (MOCA) Database: A Census of Nearby Associations, Open Clusters, and Young Substellar Objects within 500 pc of the Sun}",
      journal = {arXiv e-prints},
         year = 2026,
        month = feb,
          eid = {arXiv:2602.15695},
        pages = {arXiv:2602.15695},
archivePrefix = {arXiv},
       eprint = {2602.15695},
 primaryClass = {astro-ph.SR},
       adsurl = {https://ui.adsabs.harvard.edu/abs/2026arXiv260215695G}
}

@article{morley2024sonora,
  title={The Sonora substellar atmosphere models. III. Diamondback: atmospheric properties, spectra, and evolution for warm cloudy substellar objects},
  author={Morley, Caroline V and Mukherjee, Sagnick and Marley, Mark S and Fortney, Jonathan J and Visscher, Channon and Lupu, Roxana and Gharib-Nezhad, Ehsan and Thorngren, Daniel and Freedman, Richard and Batalha, Natasha},
  journal={The Astrophysical Journal},
  volume={975},
  number={1},
  pages={59},
  year={2024},
  publisher={The American Astronomical Society}
}

@article{saumon2008evolution,
  title={The evolution of L and T dwarfs in color-magnitude diagrams},
  author={Saumon, D and Marley, Mark S},
  journal={The Astrophysical Journal},
  volume={689},
  number={2},
  pages={1327--1344},
  year={2008}
}

@article{burningham2021cloud,
  title={Cloud busting: enstatite and quartz clouds in the atmosphere of 2M2224-0158},
  author={Burningham, Ben and Faherty, Jacqueline K and Gonzales, Eileen C and Marley, Mark S and Visscher, Channon and Lupu, Roxana and Gaarn, Josefine and Fabienne Bieger, Michelle and Freedman, Richard and Saumon, Didier},
  journal={Monthly Notices of the Royal Astronomical Society},
  volume={506},
  number={2},
  pages={1944--1961},
  year={2021},
  publisher={Oxford University Press}
}

@article{yang2016extrasolar,
  title={Extrasolar storms: pressure-dependent changes in light-curve phase in brown dwarfs from simultaneous HST and Spitzer observations},
  author={Yang, Hao and Apai, D{\'a}niel and Marley, Mark S and Karalidi, Theodora and Flateau, Davin and Showman, Adam P and Metchev, Stanimir and Buenzli, Esther and Radigan, Jacqueline and Artigau, {\'E}tienne and others},
  journal={The Astrophysical Journal},
  volume={826},
  number={1},
  pages={8},
  year={2016},
  publisher={The American Astronomical Society}
}

@article{plummer2026mapping,
  title={Mapping the Cloud-driven Atmospheric Dynamics and Chemistry of an Isolated Exoplanet Analog with Harmonic Signatures},
  author={Plummer, Michael K and Cocchini, Francis P and Kearns, Peter A and McCarthy, Allison M and Artigau, {\'E}tienne and Cowan, Nicolas B and Akhmetshyn, Roman and Vos, Johanna M and Nasedkin, Evert and Visscher, Channon and others},
  journal={The Astronomical Journal},
  volume={171},
  number={3},
  pages={195},
  year={2026},
  publisher={The American Astronomical Society}
}

@article{morley2014water,
  title={Water clouds in Y dwarfs and exoplanets},
  author={Morley, Caroline V and Marley, Mark S and Fortney, Jonathan J and Lupu, Roxana and Saumon, Didier and Greene, Tom and Lodders, Katharina},
  journal={The Astrophysical Journal},
  volume={787},
  number={1},
  pages={78},
  year={2014},
  publisher={The American Astronomical Society}
}

@article{tannock2021weather,
  title={Weather on other worlds. V. the three most rapidly rotating Ultra-cool Dwarfs},
  author={Tannock, Megan E and Metchev, Stanimir and Heinze, Aren and Miles-P{\'a}ez, Paulo A and Gagn{\'e}, Jonathan and Burgasser, Adam and Marley, Mark S and Apai, D{\'a}niel and Su{\'a}rez, Genaro and Plavchan, Peter},
  journal={The Astronomical Journal},
  volume={161},
  number={5},
  pages={224},
  year={2021},
  publisher={The American Astronomical Society}
}

@article{scholz2015rotation,
  title={Rotation periods of young brown dwarfs: K2 survey in Upper Scorpius},
  author={Scholz, Alexander and Kostov, Veselin and Jayawardhana, Ray and Mu{\v{z}}i{\'c}, Koraljka},
  journal={The Astrophysical Journal Letters},
  volume={809},
  number={2},
  pages={L29},
  year={2015},
  publisher={The American Astronomical Society}
}

@article{moore2019rotation,
  title={The rotation-disk connection in young brown dwarfs: strong evidence for early rotational braking},
  author={Moore, Keavin and Scholz, Aleks and Jayawardhana, Ray},
  journal={The Astrophysical Journal},
  volume={872},
  number={2},
  pages={159},
  year={2019},
  publisher={The American Astronomical Society}
}

@article{radigan2012large,
  title={Large-amplitude variations of an L/T transition brown dwarf: multi-wavelength observations of patchy, high-contrast cloud features},
  author={Radigan, Jacqueline and Jayawardhana, Ray and Lafreni{\`e}re, David and Artigau, Etienne and Marley, Mark and Saumon, Didier},
  journal={The Astrophysical Journal},
  volume={750},
  number={2},
  pages={105},
  year={2012},
  publisher={The American Astronomical Society}
}

@article{dupuy2012hawaii,
  title={The Hawaii infrared parallax program. I. Ultracool binaries and the L/T transition},
  author={Dupuy, Trent J and Liu, Michael C},
  journal={The Astrophysical Journal Supplement Series},
  volume={201},
  number={2},
  pages={19},
  year={2012},
  publisher={The American Astronomical Society}
}

@article{dupuy2013distances,
  title={Distances, luminosities, and temperatures of the coldest known substellar objects},
  author={Dupuy, Trent J and Kraus, Adam L},
  journal={Science},
  volume={341},
  number={6153},
  pages={1492--1495},
  year={2013},
  publisher={American Association for the Advancement of Science}
}

@article{deacon2014wide,
  title={Wide cool and ultracool companions to nearby stars from Pan-STARRS 1},
  author={Deacon, Niall R and Liu, Michael C and Magnier, Eugene A and Aller, Kimberly M and Best, William MJ and Dupuy, Trent and Bowler, Brendan P and Mann, Andrew W and Redstone, Joshua A and Burgett, William S and others},
  journal={The Astrophysical Journal},
  volume={792},
  number={2},
  pages={119},
  year={2014},
  publisher={The American Astronomical Society}
}

@article{liu2016hawaii,
  title={The Hawaii infrared parallax program. II. Young ultracool field dwarfs},
  author={Liu, Michael C and Dupuy, Trent J and Allers, Katelyn N},
  journal={The Astrophysical Journal},
  volume={833},
  number={1},
  pages={96},
  year={2016},
  publisher={The American Astronomical Society}
}

@article{best2018photometry,
  title={Photometry and Proper Motions of M, L, and T Dwarfs from the Pan-STARRS1 3 $\pi$ Survey},
  author={Best, William MJ and Magnier, Eugene A and Liu, Michael C and Aller, Kimberly M and Zhang, Zhoujian and Burgett, WS and Chambers, KC and Draper, P and Flewelling, H and Kaiser, N and others},
  journal={The Astrophysical Journal Supplement Series},
  volume={234},
  number={1},
  pages={1},
  year={2018},
  publisher={The American Astronomical Society}
}

@article{best2021volume,
  title={A volume-limited sample of ultracool dwarfs. I. Construction, space density, and a gap in the L/T transition},
  author={Best, William MJ and Liu, Michael C and Magnier, Eugene A and Dupuy, Trent J},
  journal={The Astronomical Journal},
  volume={161},
  number={1},
  pages={42},
  year={2021},
  publisher={The American Astronomical Society}
}

@article{sanghi2023hawaii,
  title={The Hawaii Infrared Parallax Program. VI. The fundamental properties of 1000+ ultracool dwarfs and planetary-mass objects using optical to mid-infrared spectral energy distributions and comparison to BT-Settl and ATMO 2020 model atmospheres},
  author={Sanghi, Aniket and Liu, Michael C and Best, William MJ and Dupuy, Trent J and Siverd, Robert J and Zhang, Zhoujian and Hurt, Spencer A and Magnier, Eugene A and Aller, Kimberly M and Deacon, Niall R},
  journal={The Astrophysical Journal},
  volume={959},
  number={1},
  pages={63},
  year={2023},
  publisher={The American Astronomical Society}
}

@article{artigau2018variability,
  title={Variability of brown dwarfs},
  author={Artigau, {\'E}tienne},
  journal={arXiv preprint arXiv:1803.07672},
  year={2018}
}

@article{vos2017viewing,
  title={The viewing geometry of brown dwarfs influences their observed colors and variability amplitudes},
  author={Vos, Johanna M and Allers, Katelyn N and Biller, Beth A},
  journal={The Astrophysical Journal},
  volume={842},
  number={2},
  pages={78},
  year={2017},
  publisher={The American Astronomical Society}
}

@article{burrows1997nongray,
  title={A nongray theory of extrasolar giant planets and brown dwarfs},
  author={Burrows, A and Marley, M and Hubbard, WB and Lunine, JI and Guillot, T and Saumon, D and Freedman, R and Sudarsky, D and Sharp, C},
  journal={The Astrophysical Journal},
  volume={491},
  number={2},
  pages={856--875},
  year={1997}
}

@article{miles2017rotation,
  title={Rotation periods and photometric variability of rapidly rotating ultracool dwarfs},
  author={Miles-P{\'a}ez, PA and Pall{\'e}, E and Zapatero Osorio, MR},
  journal={Monthly Notices of the Royal Astronomical Society},
  volume={472},
  number={2},
  pages={2297--2314},
  year={2017},
  publisher={Oxford University Press}
}

@ARTICLE{starry_p,
       author = {{Luger}, Rodrigo and {Foreman-Mackey}, Daniel and {Hedges}, Christina},
        title = "{starry\_process: Interpretable Gaussian processes for stellar light curves}",
      journal = {The Journal of Open Source Software},
         year = 2021,
        month = jul,
       volume = {6},
       number = {63},
          eid = {3071},
        pages = {3071},
          doi = {10.21105/joss.03071},
archivePrefix = {arXiv},
       eprint = {2102.01774},
 primaryClass = {astro-ph.SR},
       adsurl = {https://ui.adsabs.harvard.edu/abs/2021JOSS....6.3071L}
}

@article{cowan2025maximizing,
  title={Maximizing Ariel's Survey Leverage for Population-Level Studies of Exoplanets},
  author={Cowan, Nicolas B and Coull-Neveu, Ben},
  journal={arXiv preprint arXiv:2506.06429},
  year={2025}
}

@article{panek2026balancing,
  title={Balancing Variety and Sample Size: Optimal Parameter Sampling for Ariel Target Selection},
  author={Panek, Emilie and Roman, Alexander and Matcheva, Katia and Matchev, Konstantin T and Cowan, Nicolas B},
  journal={arXiv preprint arXiv:2601.21020},
  year={2026}
}

@article{radigan2014independent,
  title={An independent analysis of the brown dwarf atmosphere monitoring (BAM) data: large-amplitude variability is rare outside the L/T transition},
  author={Radigan, Jacqueline},
  journal={The Astrophysical Journal},
  volume={797},
  number={2},
  pages={120},
  year={2014},
  publisher={The American Astronomical Society}
}

@article{lew2016cloud,
  title={Cloud atlas: discovery of patchy clouds and high-amplitude rotational modulations in a young, extremely red L-type brown dwarf},
  author={Lew, Ben WP and Apai, Daniel and Zhou, Yifan and Schneider, Glenn and Burgasser, Adam J and Karalidi, Theodora and Yang, Hao and Marley, Mark S and Cowan, Nicolas B and Bedin, Luigi R and others},
  journal={The Astrophysical Journal Letters},
  volume={829},
  number={2},
  pages={L32},
  year={2016},
  publisher={The American Astronomical Society}
}

@article{rauscher2007hot,
  title={Hot Jupiter variability in eclipse depth},
  author={Rauscher, Emily and Menou, Kristen and Cho, James Y-K and Seager, Sara and Hansen, Bradley MS},
  journal={The Astrophysical Journal Letters},
  volume={662},
  number={2},
  pages={L115--L118},
  year={2007}
}

@article{komacek2020temporal,
  title={Temporal variability in hot Jupiter atmospheres},
  author={Komacek, Thaddeus D and Showman, Adam P},
  journal={The Astrophysical Journal},
  volume={888},
  number={1},
  pages={2},
  year={2020},
  publisher={The American Astronomical Society}
}

@article{cho2021storms,
  title={Storms, variability, and multiple equilibria on hot Jupiters},
  author={Cho, James YK and Skinner, Jack W and Thrastarson, Heidar Th},
  journal={The Astrophysical Journal Letters},
  volume={913},
  number={2},
  pages={L32},
  year={2021},
  publisher={The American Astronomical Society}
}

@article{agol2010climate,
  title={The climate of HD 189733b from fourteen transits and eclipses measured by Spitzer},
  author={Agol, Eric and Cowan, Nicolas B and Knutson, Heather A and Deming, Drake and Steffen, Jason H and Henry, Gregory W and Charbonneau, David},
  journal={The Astrophysical Journal},
  volume={721},
  number={2},
  pages={1861--1877},
  year={2010},
  publisher={The American Astronomical Society}
}

@article{knutson2011spitzer,
  title={A Spitzer transmission spectrum for the exoplanet GJ 436b, evidence for stellar variability, and constraints on dayside flux variations},
  author={Knutson, Heather A and Madhusudhan, Nikku and Cowan, Nicolas B and Christiansen, Jessie L and Agol, Eric and Deming, Drake and D{\'e}sert, Jean-Michel and Charbonneau, David and Henry, Gregory W and Homeier, Derek and others},
  journal={The Astrophysical Journal},
  volume={735},
  number={1},
  pages={27},
  year={2011},
  publisher={The American Astronomical Society}
}

@article{bell2019mass,
  title={Mass loss from the exoplanet WASP-12b inferred from Spitzer phase curves},
  author={Bell, Taylor J and Zhang, Michael and Cubillos, Patricio E and Dang, Lisa and Fossati, Luca and Todorov, Kamen O and Cowan, Nicolas B and Deming, Drake and Zellem, Robert T and Stevenson, Kevin B and others},
  journal={Monthly Notices of the Royal Astronomical Society},
  volume={489},
  number={2},
  pages={1995--2013},
  year={2019},
  publisher={Oxford University Press}
}

@article{sutlieff2023measuring,
  title={Measuring the variability of directly imaged exoplanets using vector Apodizing Phase Plates combined with ground-based differential spectrophotometry},
  author={Sutlieff, Ben J and Birkby, Jayne L and Stone, Jordan M and Doelman, David S and Kenworthy, Matthew A and Panwar, Vatsal and Bohn, Alexander J and Ertel, Steve and Snik, Frans and Woodward, Charles E and others},
  journal={Monthly Notices of the Royal Astronomical Society},
  volume={520},
  number={3},
  pages={4235--4257},
  year={2023},
  publisher={Oxford University Press}
}

@article{kostov2013mapping,
  title={Mapping directly imaged giant exoplanets},
  author={Kostov, Veselin and Apai, D{\'a}niel},
  journal={The Astrophysical Journal},
  volume={762},
  number={1},
  pages={47},
  year={2013},
  publisher={The American Astronomical Society}
}

@article{biller2018exoplanet,
  title={Exoplanet atmosphere measurements from direct imaging},
  author={Biller, Beth A and Bonnefoy, Micka{\~A}{\c{G}}l},
  journal={arXiv preprint arXiv:1807.05136},
  year={2018}
}

@article{changeat2025synergetic,
  title={On the synergetic use of Ariel and JWST for exoplanet atmospheric science},
  author={Changeat, Quentin and Lagage, Pierre-Olivier and Tinetti, Giovanna and Charnay, Benjamin and Cowan, Nicolas B and Danielski, Camilla and Ducrot, Elsa and Dyrek, Achrene and Edwards, Billy and Ikoma, Masahiro and others},
  journal={arXiv preprint arXiv:2509.02657},
  year={2025}
}

@article{croll2016long,
  title={Long-term, Multiwavelength Light Curves of Ultra-Cool Dwarfs: II. The evolving Light Curves of the T2. 5 SIMP 0136 \& the Uncorrelated Light Curves of the M9 TVLM 513},
  author={Croll, Bryce and Muirhead, Philip S and Lichtman, Jack and Han, Eunkyu and Dalba, Paul A and Radigan, Jacqueline},
  journal={arXiv preprint arXiv:1609.03587},
  year={2016}
}

@article{baraffe2015new,
  title={New evolutionary models for pre-main sequence and main sequence low-mass stars down to the hydrogen-burning limit},
  author={Baraffe, Isabelle and Homeier, Derek and Allard, France and Chabrier, Gilles},
  journal={Astronomy \& Astrophysics},
  volume={577},
  pages={A42},
  year={2015},
  publisher={EDP Sciences}
}

@article{zhang2022vizier,
  title={VizieR Online Data Catalog: New L subdwarfs, population properties (Zhang+, 2018)},
  author={Zhang, ZH and Galvez-Ortiz, MC and Pinfield, DJ and Burgasser, AJ and Lodieu, N and Jones, HRA and Martin, EL and Burningham, B and Homeier, D and Allard, F and others},
  journal={VizieR Online Data Catalog},
  volume={748},
  pages={J--MNRAS},
  year={2022}
}

@article{reid2009vizier,
  title={VizieR Online Data Catalog: Ultracool dwarfs from the 2MASS (Reid+, 2008)},
  author={Reid, NI and Cruz, KL and Kirkpatrick, JD and Allen, PR and Mungall, F and Liebert, J and Lowrance, P and Sweet, A},
  journal={VizieR Online Data Catalog},
  volume={513},
  pages={J--AJ},
  year={2009}
}

@article{salim2003lsr,
  title={LSR 0602+ 3910: Discovery of a bright nearby L-type brown dwarf},
  author={Salim, Samir and L{\'e}pine, S{\'e}bastien and Rich, R Michael and Shara, Michael M},
  journal={The Astrophysical Journal Letters},
  volume={586},
  number={2},
  pages={L149--L152},
  year={2003}
}

@article{corbally2005commission,
  title={Commission 45: Stellar Classification},
  author={Corbally, Christopher and Giridhar, Sunetra and Bailer-Jones, Coryn and Humphreys, Roberta and Kirkpatrick, Davy and Evans, Tom Lloyd and Luri, Xavier and Minniti, Dante and Pasinetti, Laura and Strai{\v{z}}ys, Vytautas and others},
  journal={Proceedings of the International Astronomical Union},
  volume={1},
  number={T26A},
  pages={221--231},
  year={2005},
  publisher={Cambridge University Press}
}

@article{lodieu2005vizier,
  title={VizieR Online Data Catalog: Southern red high proper motion objects (Lodieu+, 2005)},
  author={Lodieu, N and Scholz, R-D and McCaughrean, MJ and Ibata, R and Irwin, M and Zinnecker, H},
  journal={VizieR Online Data Catalog},
  volume={344},
  pages={J--A+},
  year={2005}
}

@article{baron2015discovery,
  title={Discovery and characterization of wide binary systems with a very low mass component},
  author={Baron, Fr{\'e}d{\'e}rique and Lafreni{\`e}re, David and Artigau, {\'E}tienne and Doyon, Ren{\'e} and Gagn{\'e}, Jonathan and Davison, Cassy L and Malo, Lison and Robert, Jasmin and Nadeau, Daniel and Reyl{\'e}, C{\'e}line},
  journal={The Astrophysical Journal},
  volume={802},
  number={1},
  pages={37},
  year={2015},
  publisher={The American Astronomical Society}
}

@article{kirkpatrick1999dwarfs,
  title={Dwarfs cooler than “M”: The definition of spectral type “L” using discoveries from the 2-Micron All-Sky Survey (2MASS)},
  author={Kirkpatrick, J Davy and Reid, I Neill and Liebert, James and Cutri, Roc M and Nelson, Brant and Beichman, Charles A and Dahn, Conard C and Monet, David G and Gizis, John E and Skrutskie, Michael F},
  journal={The Astrophysical Journal},
  volume={519},
  number={2},
  pages={802--833},
  year={1999}
}

@article{kirkpatrick200067,
  title={67 additional L dwarfs discovered by the Two Micron All Sky Survey},
  author={Kirkpatrick, J Davy and Reid, I Neill and Liebert, James and Gizis, John E and Burgasser, Adam J and Monet, David G and Dahn, Conard C and Nelson, Brant and Williams, Rik J},
  journal={The Astronomical Journal},
  volume={120},
  number={1},
  pages={447--472},
  year={2000}
}

@article{schneider2023astrometry,
  title={Astrometry and photometry for 1000 L, T, and Y dwarfs from the UKIRT Hemisphere Survey},
  author={Schneider, Adam C and Munn, Jeffrey A and Vrba, Frederick J and Bruursema, Justice and Dahm, Scott E and Williams, Stephen J and Liu, Michael C and Dorland, Bryan N},
  journal={The Astronomical Journal},
  volume={166},
  number={3},
  pages={103},
  year={2023},
  publisher={The American Astronomical Society}
}

@article{gagne2018banyan,
  title={BANYAN. XI. The BANYAN $\Sigma$ multivariate Bayesian algorithm to identify members of young associations with 150 pc},
  author={Gagn{\'e}, Jonathan and Mamajek, Eric E and Malo, Lison and Riedel, Adric and Rodriguez, David and Lafreni{\`e}re, David and Faherty, Jacqueline K and Roy-Loubier, Olivier and Pueyo, Laurent and Robin, Annie C and others},
  journal={The Astrophysical Journal},
  volume={856},
  number={1},
  pages={23},
  year={2018},
  publisher={The American Astronomical Society}
}

@article{allers2013near,
  title={A near-infrared spectroscopic study of young field ultracool dwarfs},
  author={Allers, Katelyn N and Liu, Michael C},
  journal={The Astrophysical Journal},
  volume={772},
  number={2},
  pages={79},
  year={2013},
  publisher={The American Astronomical Society}
}

@article{koen2004band,
  title={I-band time-series observations of five bright ultracool dwarfs},
  author={Koen, Chris},
  journal={Monthly Notices of the Royal Astronomical Society},
  volume={354},
  number={2},
  pages={378--386},
  year={2004},
  publisher={Blackwell Science Ltd}
}

@article{lambier2025rotation,
  title={Rotation Periods of Candidate Single Late-M Dwarfs in TESS},
  author={Lambier, Samantha and Metchev, Stanimir and Miles-P{\'a}ez, Paulo and Moranta, Leslie and Wolfe, Dakota and Hales, Joelene and Martinovic, Jeffrey},
  journal={The Astronomical Journal},
  volume={170},
  number={3},
  pages={168},
  year={2025},
  publisher={The American Astronomical Society}
}

% Alternatively you could enter them by hand, like this:
% This method is tedious and prone to error if you have lots of references
%\begin{thebibliography}{99}
%\bibitem[\protect\citeauthoryear{Author}{2012}]{Author2012}
%Author A.~N., 2013, Journal of Improbable Astronomy, 1, 1
%\bibitem[\protect\citeauthoryear{Others}{2013}]{Others2013}
%Others S., 2012, Journal of Interesting Stuff, 17, 198
%\end{thebibliography}

%%%%%%%%%%%%%%%%%%%%%%%%%%%%%%%%%%%%%%%%%%%%%%%%%%

%%%%%%%%%%%%%%%%% APPENDICES %%%%%%%%%%%%%%%%%%%%%
\onecolumn
\appendix

\section{The table of brightest targets}
\label{sec:appendix1}
In this section we present a reduced Table~\ref{tab:best} with the known physical properties of the brightest isolated brown dwarfs that could potentially be guided on with Ariel/FGS2 at 1 Hz cadence. 

\begin{table*}
\centering
\begin{threeparttable}
\caption{Brightest isolated L dwarfs from our sample, sorted by the photon flux in the FGS2 band. The SNR of each spectroscopic instrument is the median value across whole wavelength range. Physical properties such as mass and log(g) are derived from
the evolutionary models given the age and bolometric luminosity, the former is known if the object is in
the stellar association. Hence field dwarfs do not have those measurements. The full table is available online.}
\label{tab:best}
\begin{tabular}{lccccccccccc}
\toprule
Designation & \makecell{RA \\ (deg.)} & \makecell{Dec \\ (deg.)} & Sp.T. &
\makecell{Age \\ (Myr)} & \makecell{Mass \\ ($M_{\rm J}$)} &
\makecell{Log(g)} & \makecell{Period \\ (h)} & \makecell{FGS2 flux \\ ($10^3$ $\gamma$/s/m$^2$)} &
\makecell{NIRSpec \\ SNR} & \makecell{AIRS0 \\ SNR} & \makecell{AIRS1 \\ SNR} \\
\midrule
LSR J1826+3014 & 276.534 & 30.235 & L0 $^{(1)}$ & --- & --- & --- & --- & 54.8 & 1534 & 438 & 618 \\
2MASS J1731+2721 & 262.873 & 27.355 & L0 $^{(2)}$ & 149$^{+51}_{-19}$$^{(10)}$ & 44$^{+3}_{-3}$$^{(11)}$ & 4.91$^{+0.04}_{-0.04}$$^{(11)}$ & --- & 45.4 & 1440 & 436 & 625 \\

LSR J0602+3910 & 90.627 & 39.183 & L2 $^{(3)}$ & 135$^{+115}_{-115}$$^{(12)}$ & 28$^{+11}_{-11}$$^{(11)}$ & 4.62$^{+0.25}_{-0.25}$$^{(11)}$ & --- & 42.1 & 1386 & 437 & 627 \\

2MASS J1645-1319 & 251.340 & -13.334 & L1 $^{(4)}$ & --- & --- & --- & --- & 35.5 & 1273 & 401 & 576 \\

SSSPM J0829-1309 & 127.139 & -13.155 & L1 $^{(5)}$ & --- & --- & --- & 2.9 $^{(13)}$ & 31.9 & 1175 & 368 & 525 \\

2MASS J1259+1001 & 194.925 & 10.028 & L4.5 $^{(6)}$ & --- & --- & --- & --- & 30.5 & 1063 & 282 & 378 \\

2MASS J1439+1929 & 219.861 & 19.489 & L1 $^{(7)}$ & --- & --- & --- & --- & 28.6 & 1110 & 331 & 460 \\

2MASS J1555-0956 & 238.819 & -9.938 & L1 $^{(4)}$ & --- & --- & --- & --- & 27.9 & 1131 & 342 & 491 \\

2MASS J0036+1821 & 9.0713 & 18.353 & L3 $^{(8)}$ & --- & --- & --- & 2.7 $^{(14)}$ & 27.3 & 1216 & 441 & 676 \\

2MASS J1300+1912 & 195.173 & 19.204 & L1 $^{(9)}$ & --- & --- & --- & --- & 23.7 & 1042 & 315 & 452 \\
\bottomrule
\end{tabular}

\footnotesize
\textbf{References}: (1) \cite{zhang2022vizier};
(2) \cite{reid2009vizier};
(3) \cite{salim2003lsr};
(4) \cite{corbally2005commission};
(5) \cite{lodieu2005vizier};
(6) \cite{baron2015discovery};\\
(7) \cite{kirkpatrick1999dwarfs};
(8) \cite{kirkpatrick200067};
(9) \cite{schneider2023astrometry};
(10) \cite{gagne2018banyan};
(11) \cite{baraffe2015new};
(12) \cite{allers2013near};\\
(13) \cite{koen2004band};
(14) \cite{lambier2025rotation}.

\end{threeparttable}

\end{table*}

%%%%%%%%%%%%%%%%%%%%%%%%%%%%%%%%%%%%%%%%%%%%%%%%%%

% Don't change these lines
\bsp	% typesetting comment
\label{lastpage}
\end{document}